\RequirePackage{lineno}
\documentclass[superscriptaddress,amsmath,amssymb,nofootinbib]{revtex4}

\usepackage{graphicx}
\usepackage{dcolumn}
\usepackage{bm}
\usepackage{ulem}
\usepackage{xfrac}
\usepackage{lipsum}

\usepackage[printwatermark]{xwatermark}
\usepackage{xcolor}

\usepackage{graphicx,amsmath,amssymb}
\usepackage{epstopdf}
\usepackage{lineno}
\usepackage[colorlinks=true, urlcolor=blue, linkcolor=blue, citecolor=blue]{hyperref}

\begin{document}

\title{Experimental study of the behavior of the Bjorken sum at very low $Q^2$}

\author{A.~Deur} 
\affiliation{Thomas Jefferson National Accelerator Facility, Newport News, Virginia 23606, USA}
\affiliation{University of Virginia, Charlottesville, Virginia 22904, USA}
\author{J.~P.~Chen}  
\affiliation{Thomas Jefferson National Accelerator Facility, Newport News, Virginia 23606, USA}
\author{S.~E.~Kuhn} 
\affiliation{Old Dominion University,  Norfolk, Virginia 23529, USA}
\author{C.~Peng}    
\affiliation{Duke University, Durham, NC 27708, USA}
\author{M.~Ripani}  
\affiliation{INFN, Sezione di Genova, 16146 Genova, Italy}
\author{V.~Sulkosky}  
\affiliation{College of William and Mary, Williamsburg, VA 23187, USA}
\affiliation{Thomas Jefferson National Accelerator Facility, Newport News, Virginia 23606, USA}
\affiliation{University of Virginia, Charlottesville, Virginia 22904, USA}
\author{K.~Adhikari} 
\affiliation{Old Dominion University,  Norfolk, Virginia 23529, USA}
\affiliation{Thomas Jefferson National Accelerator Facility, Newport News, Virginia 23606, USA}
\affiliation{Mississippi State University, Mississippi State, Mississippi, MS 39762, USA}
\author{M.~Battaglieri}
\affiliation{Thomas Jefferson National Accelerator Facility, Newport News, Virginia 23606, USA}  
\affiliation{INFN, Sezione di Genova, 16146 Genova, Italy}
\author{V.~D.~Burkert}
\affiliation{Thomas Jefferson National Accelerator Facility, Newport News, Virginia 23606, USA}
\author{G. D. Cates}
\affiliation{University of Virginia, Charlottesville, Virginia 22904, USA}
\author{R. De Vita}
\affiliation{INFN, Sezione di Genova, 16146 Genova, Italy}
\author{G. E. Dodge}
\affiliation{Old Dominion University, Norfolk, Virginia 23529, USA}
\author{L.~El~Fassi}  
\affiliation{Old Dominion University,  Norfolk, Virginia 23529, USA}
\affiliation{Mississippi State University, Mississippi State, Mississippi, MS 39762, USA}
\author{F.  Garibaldi}
\affiliation{INFN, Sezione di Roma,  I-00185 Rome, Italy}
\affiliation{Istituto Superiore di Sanit\`a, I-00161 Rome, Italy}
\author{H.~Kang}  
\affiliation{Seoul National University, Seoul, Korea}
\author{M.~Osipenko}   
\affiliation{INFN, Sezione di Genova, 16146 Genova, Italy}
\author{J.~T.~Singh}  
\affiliation{University of Virginia, Charlottesville, Virginia 22904, USA}
\author{K.~Slifer}  
\affiliation{University of Virginia, Charlottesville, Virginia 22904, USA}
\affiliation{University of New Hampshire, Durham, NH 03824, USA}
\author{J.~Zhang}  
\affiliation{University of Virginia, Charlottesville, Virginia 22904, USA}
\author{X.~Zheng}  
\affiliation{University of Virginia, Charlottesville, Virginia 22904, USA}


\begin{abstract}

We present new data on the Bjorken sum $\overline \Gamma_1^{p-n}(Q^2)$ at 
4-momentum transfer $ 0.021 \leq Q^2 \leq 0.496$ GeV$^2$. The data were 
obtained in two experiments performed at Jefferson Lab: 
EG4 on polarized protons and deuterons, and E97110 on polarized $^3$He from which neutron data were extracted. 
The data cover the domain where chiral effective field theory ($\chi$EFT), 
the leading effective theory of the Strong Force at large distances, is expected to be applicable. 
We find that our data and the predictions from $\chi$EFT are only in marginal agreement. 
This is somewhat surprising as the contribution from the $\Delta(1232)$ resonance is suppressed 
in this observable, which should make it more reliably predicted by $\chi$EFT
 than quantities in which the $\Delta$ contribution is important.
The data are also compared to a number of phenomenological models with various degrees of agreement.

\end{abstract}


\maketitle


The archetype of spin sum rules, the Bjorken sum rule~\cite{Bjorken:1966jh}, has 
played a central role in the investigation of nucleon spin 
structure~\cite{Deur:2018roz}. 
The sum rule stands at infinite $Q^2$, the squared four-momentum transferred between the 
probing beam and the probed nucleon, and
relates the nucleon flavor-singlet axial charge 
$g_{A}$ to the isovector part of the integrated spin-dependent structure function $g_1(x)$:
\begin{linenomath}
\begin{equation}
\overline\Gamma_1^{p-n} \equiv 
\overline\Gamma_1^{p}-\overline\Gamma_1^{n}\equiv\int_{0}^{1^-} \big[g_1^{p}(x)-g_1^{n}(x)\big] dx =\frac{g_{A}}{6}.
\label{eq:bj}
\end{equation}
\end{linenomath}
Here, $x\equiv Q^2/(2M\nu)$ is Bjorken scaling variable, $M$ the nucleon mass  
and $\nu$ the energy transfer between the incoming lepton and the nucleon. 
$g_1^{p(n)}(x)$ denotes the proton (neutron) quantity.
The bars over $\Gamma_1$ and the $1^-$ integral limit indicate that the elastic contribution is excluded.
The value of the axial charge is measured  independently {\it via} neutron $\beta$-decay, $g_A=1.2762(5)$~\cite{PDG2020}.
Measurements of $\overline\Gamma_1$, performed at SLAC~\cite{Abe:1994cp, Anthony:1996mw}, 
CERN~\cite{Adeva:1993km}, DESY~\cite{Ackerstaff:1997ws} and Jefferson Lab (JLab)~\cite{Deur:2004ti,Deur:2008ej,Deur:2014vea},
by scattering polarized leptons off polarized targets,
are at finite $Q^2$ values. In that case, $g_1(x)$ and $\overline \Gamma_1$ acquire a $Q^{2}$-dependence, 
which is calculable at $Q^2 \gtrsim 1$ GeV$^2$ with perturbative quantum chromodynamics 
(pQCD)~\cite{Kataev:1994gd}, and at $Q^2 \ll 1$ GeV$^2$ with chiral effective field theory
($\chi$EFT)~\cite{Bernard:1992nz, Bernard:2012hb,Lensky:2014dda,Alarcon:2020icz}, an effective theory of QCD~\cite{Bernard:1995dp}.
At $Q^2 \to 0$, $\overline\Gamma_1$ relates to the Gerasimov-Drell-Hearn (GDH)
sum rule~\cite{Gerasimov:1965et}, which has been verified   for the proton within experimental uncertainty~\cite{Dutz:2003mm}. 
The GDH sum rule predicts:
\begin{linenomath}
\begin{equation}
\overline\Gamma_1^{p-n}(Q^2)|_{_{Q^2  \to 0}}  = \frac{Q^2}{8}\bigg(\frac{\kappa_n^2}{M_n^2}-\frac{\kappa_p^2}{M_p^2}\bigg),
\label{eq:bj-GDH}
\end{equation}
\end{linenomath}
where $\kappa_n$ and $\kappa_p$ are,  respectively, the anomalous magnetic moments of the neutron and proton~\cite{PDG2020}.
Since $\kappa_n^2/M_n^2 > \kappa_p^2/M_p^2$, $\overline \Gamma_1^{p-n}(Q^2)$ is expected to depart from zero
with a positive slope.
Eq.~(\ref{eq:bj-GDH}) is assumed in the $\overline \Gamma_1^{p-n}(Q^2)$ calculations from $\chi$EFT  which predict the  $Q^2$-dependence of $\overline \Gamma_1^{p-n}(Q^2)$ at low $Q^2$. 

The isovector structure of $\overline \Gamma_1^{p-n}$ simplifies its theoretical calculation compared to  $\overline \Gamma_1^{p}$ or $\overline \Gamma_1^{n}$.
In particular, the suppression of the contribution of the  $\Delta(1232)~3/2^+$ excitation 
should make the $\chi$EFT prediction of $\overline\Gamma_1^{p-n}(Q^2)$  more reliable~\cite{Burkert:2000qm, Alarcon:2020icz}. While this 
expectation is consistent with early  $\overline \Gamma_1^{p-n}$ data~\cite{Deur:2004ti, Deur:2008ej, Deur:2014vea}, measurements of another observable in which the 
$\Delta$ is suppressed,  namely the  Longitudinal-Transverse interference polarizability $\delta_{LT}(Q^2)$~\cite{Amarian:2004yf},
showed that here the argument fails. This perplexing outcome triggered both 
new  experiments at JLab designed to cover well the $\chi$EFT 
domain~\cite{Adhikari:2017wox, Sulkosky:2019zmn, Zheng:2021yrn, Sulkosky:2021qmh, g2p in Hall A}, 
and improved $\chi$EFT calculations~\cite{Bernard:2012hb, Lensky:2014dda,Alarcon:2020icz} that
explicitly include the $\Delta$ by computing the $\pi-\Delta$ graphs, in contrast with the 
earlier calculations~\cite{Bernard:1992nz}.

In this article, we present new JLab data on the Bjorken sum $\overline\Gamma_1^{p-n}(Q^2)$  
for $ 0.021 \leq Q^2 \leq 0.496$ GeV$^2$ where $\chi$EFT can be   tested well. 
The data are from the experiments EG4 (polarized proton and deuteron targets, henceforth called ``EG4'', or 
``proton'' and ``deuteron'') and E97110 (polarized  $^3$He target, henceforth called ``E97110'' or ``$^3$He''). 
The experimental and analysis descriptions, including the extraction of the individual integrals $\overline\Gamma_1^{p,n,d}$,
are reported in Refs.~\cite{Adhikari:2017wox, Sulkosky:2019zmn, Zheng:2021yrn}.
To reach the $x=0$ limit of integral (\ref{eq:bj}) requires infinite energy. 
The integrals reported in Refs.~\cite{Adhikari:2017wox, Sulkosky:2019zmn, Zheng:2021yrn} cover the range down to $x =10^{-3}$, with 
the lower $x$ contributions to $\overline \Gamma_1^p$, $\overline \Gamma_1^d$ and $\overline \Gamma_1^{n}$
estimated using a parameterization of  previous data~\cite{Fersch:2017qrq}. 
This parametrization is based on a fit of the spin asymmetries $A_1$ and $A_2$ as well as the 
unpolarized structure function $F_1$ to all existing data. From these, the spin structure function $g_1$ 
is formed and integrated separately for each of the three different targets (p, d, n) up to the lowest 
measured $x$-point for each target and each $Q^2$.

 \begin{table}
\resizebox{0.47\textwidth}{!}{%
\begin{tabular}{|c|c|c|c|c|c|} \hline
$Q^2$ & $\overline\Gamma_1^{p}-\overline\Gamma_1^{n}$  (meas.) & $\overline\Gamma_1^{p-n}$ (full) & Stat. & Syst.  \\  \hline 
0.021& 0.0042  & 0.0052  & $\pm 0.0029$   &  $\pm 0.0012$   \\
0.024& 0.0008  & 0.0008  & $\pm 0.0031$   &  $\pm 0.0013$   \\ 
0.029& 0.0122  & 0.0126  & $\pm 0.0031$   &  $\pm 0.0015$   \\
0.035& 0.0031  & 0.0040  & $\pm 0.0030$   &  $\pm 0.0016$   \\
0.042& 0.0046  & 0.0080  & $\pm 0.0033$   &  $\pm 0.0016$   \\
0.050& 0.0051  & 0.0095  & $\pm 0.0035$   &  $\pm 0.0018$   \\
0.059& 0.0050 & 0.0103  & $\pm 0.0037$   &  $\pm 0.0021$   \\
0.071& 0.0065  & 0.0131  & $\pm 0.0041$   &  $\pm 0.0023$   \\
0.084& 0.0026  & 0.0107  & $\pm 0.0044$   &  $\pm 0.0023$   \\
0.101& 0.0022  & 0.0122  & $\pm 0.0042$   &  $\pm 0.0026$   \\
0.120& 0.0156  & 0.0276  & $\pm 0.0047$   &  $\pm 0.0030$   \\
0.144& 0.0051  & 0.0194  & $\pm 0.0052$   &  $\pm 0.0027$   \\
0.173& 0.0109  & 0.0266  & $\pm 0.0051$   &  $\pm 0.0027$   \\
0.205& 0.0055  & 0.0244  & $\pm 0.0061$   &  $\pm 0.0029$   \\
0.244& 0.0236  & 0.0464  & $\pm 0.0063$   &  $\pm 0.0029$   \\
0.292& 0.0230  & 0.0500  & $\pm 0.0062$   &  $\pm 0.0026$   \\
0.348& 0.0226  & 0.0533  & $\pm 0.0068$   &  $\pm 0.0027$   \\
0.416& 0.0123  & 0.0477  & $\pm 0.0073$   &  $\pm 0.0029$   \\
0.496& 0.0384  &  0.0770  & $\pm 0.0088$   &  $\pm 0.0070$   \\
\hline \hline
0.035  &   0.0062    & 0.0085	  & $\pm 0.0006$  &  $\pm  0.0018$   \\
0.057  &   0.0068    &  0.0114      &$\pm 0.0010$  &  $\pm  0.0024$   \\
0.079  &   0.0064    &  0.0128     &$\pm 0.0014$  &  $\pm  0.0028$   \\
0.101  &   0.0074    &  0.0143     &$\pm 0.0011$  &  $\pm  0.0032$   \\
0.150  &   0.0124    &  0.0220     &$\pm 0.0014$  &  $\pm  0.0051$   \\
0.200  &   0.0169    & 0.0294	 &$\pm 0.0017$   &  $\pm  0.0058$   \\
0.240  &   0.0209    & 0.0381	 & $\pm 0.0021$  &  $\pm  0.0035$   \\
\hline 	
\end{tabular}}
\vspace{-0.2cm}
\caption{Data from EG4 (top) and EG4/E97110 (bottom).
The columns show from left to right, respectively:
$Q^2$ value in GeV$^2$;
truncated integrals for $\Gamma_1^p - \Gamma_1^n$ evaluated over the $x$-ranges measured 
by the respective experiments (ranges are different for p, d and n integrals and for different $Q^2$);
full $\overline\Gamma_1^{p-n}$ after adding the estimated unmeasured low-$x$ contributions down to $x = 0.001$;
statistical uncertainty;
systematic uncertainty (including the estimate on the low-$x$ contribution).
}
\label{table_data}
\vspace{-0.7cm}
\end{table}

\begin{figure}
\protect\includegraphics[width=0.46\textwidth]{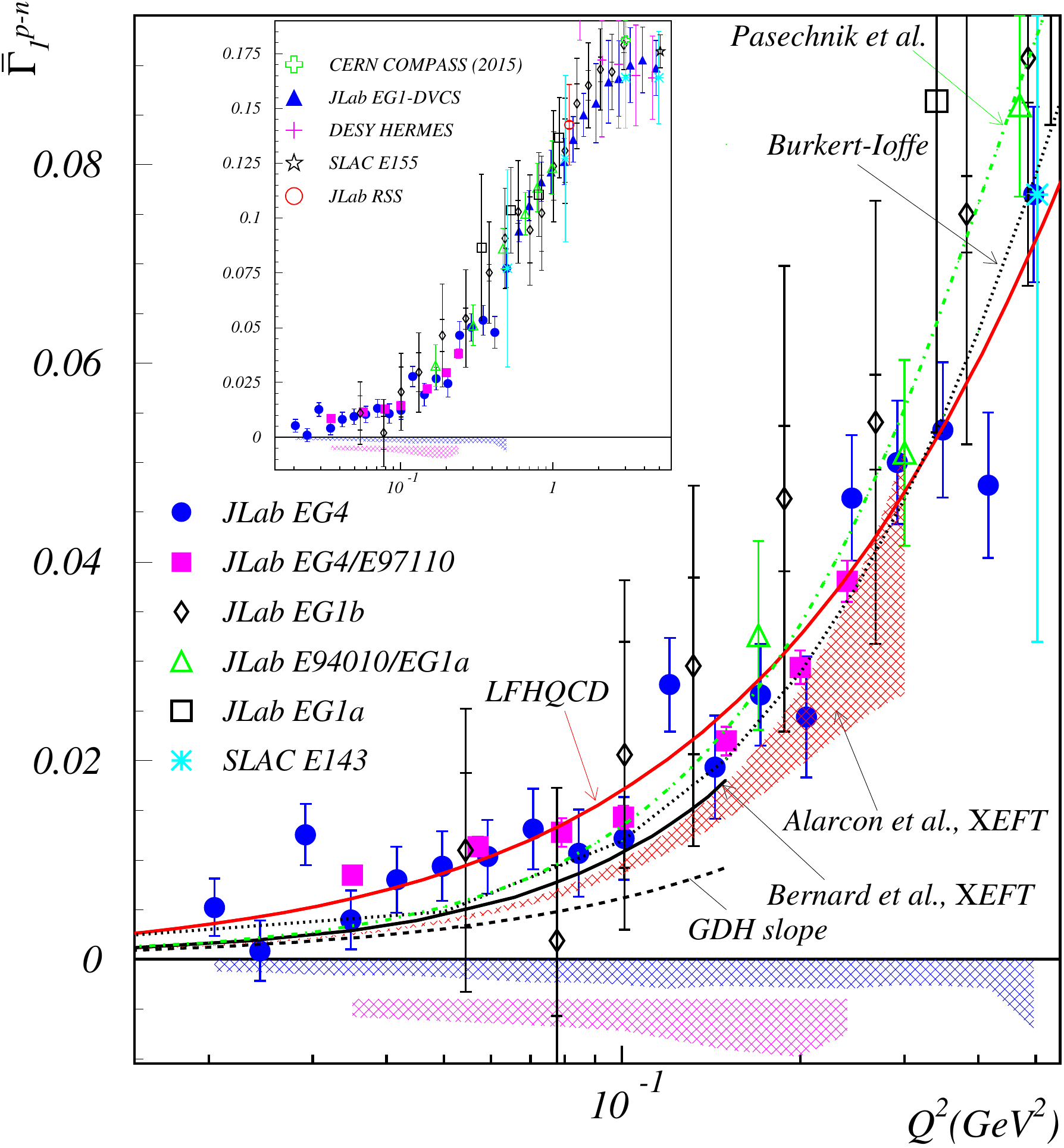}
\caption{\label{fig:bjsr}\small The Bjorken Sum $\overline\Gamma_1^{p-n}(Q^2)$ from EG4 (solid  circles), EG4/E97110  (solid squares)
and earlier data from E143 (cyan asterisk)~\cite{Abe:1994cp}, EG1a (open square)~\cite{Deur:2004ti}, E94010/EG1a (open triangles)~\cite{Deur:2004ti} and EG1b (open diamonds)~\cite{Deur:2008ej} The error bars on the EG4 and EG4/E97110 points indicate the statistical uncertainties. Their systematic uncertainties are given by the upper  blue band and lower magenta   band, respectively. The inner error bars on the other data points represent their statistical uncertainties and the outer ones the quadratic sum of the statistical and systematic uncertainties.
Also shown are the $\chi$EFT predictions from Bernard {\it et al.}~\cite{Bernard:2012hb} (black line) and
Alarc\'on {\it et al.}~\cite{Alarcon:2020icz} (red band), 
as well as model predictions~\cite{Burkert:1992tg,Pasechnik:2010fg,Brodsky:2010ur}
(see main text for details).
The embedded figure is a zoom-out to show the earlier data~\cite{Anthony:1996mw,Adeva:1993km,Ackerstaff:1997ws,Deur:2004ti,Deur:2014vea,Deur:2008ej}. 
\protect \\
}
\end{figure}

 The proton and deuteron data, analyzed at common $Q^{2}$ values, are combined as
$\overline\Gamma_1^{p-n}=2\overline\Gamma_1^{p}-\overline\Gamma_1^d/\left(1-1.5\omega_{d}\right)$
with the deuteron D-state probability $\omega_{d}=0.05\pm0.01$~\cite{omega_d} and $\overline\Gamma^d_1$
understood as  ``per nucleus". 
This accounting for the nucleon depolarization due to the deuteron D-state is the only nuclear correction
we applied, since kinematic effects like Fermi motion should not significantly affect the integrals over $x$.
We call the values obtained this way ``the EG4 data''. 
Similarly, the $n$ integrals were extracted from $^3$He by correcting for the effective polarization 
of all target nucleons~\cite{CiofidegliAtti:1996cg}.
The proton and neutron($^3$He) data were analyzed at different $Q^{2}$ values. 
Since the proton data have finer $Q^2$-bins, they were first combined into the same
number of bins as for the neutron($^3$He) data, and then linearly interpolated to the $Q^{2}$ values of the neutron($^3$He) data.
The fine binning of EG4 makes a linear extrapolation sufficient, as verified by alternatively using a quadratic interpolation and observing that the difference 
between the two interpolations is negligible compared to the experimental uncertainties.
The statistical uncertainties from the proton data were propagated according to each data point's weight in the interpolation, while the systematic uncertainties were averaged over the interpolated data points.
We call the values obtained this way ``the EG4/E97110 data''.
The two resulting (semi-independent) data sets for $\overline\Gamma_1^{p-n}$ are reported in Table~\ref{table_data} and 
shown in Fig.~\ref{fig:bjsr}, along with data from previous
experiments at larger $Q^2$~\cite{Anthony:1996mw,Adeva:1993km,Ackerstaff:1997ws,Deur:2004ti,Deur:2008ej}. 
With the new data, the world data set for $\overline\Gamma_1^{p-n}$ now spans nearly 3 orders of magnitude in $Q^2$. 
Also shown in Fig.~\ref{fig:bjsr} are the latest $\chi$EFT calculations~\cite{Bernard:2012hb,Alarcon:2020icz}
and several models. 
The Burkert-Ioffe model (dotted line) is an extrapolation of deep inelastic scattering (DIS) data based on vector
meson dominance combined with a parameterization of the resonance
contribution~\cite{Burkert:1992tg}. 
The Pasechnik {\it et al.} model~\cite{Pasechnik:2010fg} (dot-dashed line) applies analytical perturbation theory
(APT) to an earlier model
~\cite{Soffer:2004ip} that
used the smooth $Q^{2}$-dependence of $g_1+g_2$ to extrapolate DIS data to low $Q^{2}$.
 Finally, light-front holographic QCD (LFHQCD)~\cite{Brodsky:2014yha} (continuous red line) 
is a  method based on the anti-de Sitter/conformal field theory (AdS/CFT) correspondance~\cite{Maldacena:1997re} 
with QCD quantized on the light-front~\cite{Brodsky:1997de}. 
In LFHQCD, the $Q^2$-dependence of
 $\alpha_{g_1}(Q^2)$~\cite{Brodsky:2010ur} --the effective charge that folds into $\alpha_s$ the non-perturbative contributions to $\overline\Gamma_1^{p-n}$~\cite{Grunberg:1980ja, Deur:2005cf}--  is directly obtained from the AdS space curvature~\cite{Brodsky:2010ur},
a quantity uniquely determined from basic considerations e.g. that the pion mass must vanish in the chiral limit~\cite{Brodsky:2014yha, deTeramond:2014asa}.
Then, $ \overline\Gamma_1^{p-n}$ is obtained  using 
 $ \overline\Gamma_1^{p-n}=\frac{g_A}{6}(1-\frac{\alpha_{g_1}}{\pi})$. 

The $\overline\Gamma_1^{p-n}$ formed using the deuteron (EG4) or the neutron from $^3$He (E97110) agree with each other, 
indicating that for this observable, the minimal nuclear corrections we applied to obtain the neutron seem sufficient 
even at these low $Q^2$. Potential nuclear effects for deuteron and $^3$He are quite 
different: nuclear binding in $^3$He is stronger than for the deuteron, but even a small nuclear modification of the 
proton would have a much larger effect in the deuteron. 
Fig.~\ref{fig:bjsr} shows a tension between the new data and the $\chi$EFT curves.%
The two data sets display a similar tension with the models except LFHQCD~\cite{Brodsky:2010ur} with which they agree well. 
To make the above comparisons quantitative, we fit $\overline\Gamma_1^{p-n}$ 
up to $Q^2 = 0.244$ GeV$^2$, {\it viz} the domain over which the E97110 data are available.
Kinematics  impose that $\overline\Gamma_1^{p-n}(0)=0$, a constraint that we implement by using the fit function $bQ^2+cQ^4$,
with $b$ and $c$ the fit free parameters. From Eq.~(\ref{eq:bj-GDH}), the GDH sum rule predicts that 
$b=0.0618~\mathrm{GeV}^{-2} \equiv b^{\mathrm{GDH}}$. 
The Bernard {\it et al.}~\cite{Bernard:2012hb}  and Alarc\'on {\it et al.}~\cite{Alarcon:2020icz}
curves assume $b^{\mathrm{GDH}}$, and $c$ is calculated using $\chi$EFT.
The result for the best fit to the world data is given in Table~\ref{table_fit2}.
\begin{table}[t]
\resizebox{0.48\textwidth}{!}{%
\begin{tabular}{|c|c|c|} \hline
Data set                 &   $(b \pm uncor \pm cor )$      [GeV$^{-2}$]        &  $c\pm uncor \pm cor $    [GeV$^{-4}$]        \\  \hline 
World data              &  $0.182\pm0.016 \pm0.034$     &   $-0.117\pm0.091 \pm0.095$             \\  \hline 
GDH Sum Rule~\cite{Gerasimov:1965et}      &  0.0618   &     -            \\
$\chi$EFT Bernard {\it et al.}~\cite{Bernard:2012hb}  &  0.07  &   0.3        \\
$\chi$EFT Alarc\'on {\it et al.}~\cite{Alarcon:2020icz}  &   0.066(4)  &   0.25(12)                 \\
Burkert-Ioffe~\cite{Burkert:1992tg} &   0.09   &     0.3           \\
Pasechnik {\it et al.}~\cite{Pasechnik:2010fg} &   0.09  &   0.4              \\
LFHQCD~\cite{Brodsky:2010ur} &  0.177  &   -0.067              \\ \hline 	
\end{tabular}}
\vspace{-0.2cm}
\caption{Best fit of the world data 
on $ \overline\Gamma_1^{p-n}(Q^2)$   (full integral, with low-$x$ contribution)
using a fit function $bQ^2+cQ^4$. The fit is performed
up to $Q^2 = 0.244$ GeV$^2$. 
The ``$uncor$'' uncertainty designates the point-to-point 
uncorrelated uncertainty. It is the quadratic sum of the statistical uncertainty and a fraction of the systematic uncertainty determined so that $\chi^2/n.d.f = 1$ for the best fit, see Appendix. The ``$cor$'' uncertainty  is the correlated uncertainty estimated from the remaining fraction of the systematic uncertainty.
Also listed are results of fits applied to the predictions from $\chi$EFT and models. 
}
\label{table_fit2}
\vspace{-0.7cm}
\end{table}
Table~\ref{table_fit2} also shows theoretical predictions.  
For those, 
we extracted $b$ and $c$ the same way as for the data, {\it via} a fit over the region of our data.

The data points are generally above most of the theoretical calculations.
This deviation causes both the value of $c$ to   be in tension with the $\chi$EFT expectations, 
and the value of $b$ to be larger than $b^{\mathrm{GDH}}$:  
the best fit yields $b = 0.182\pm0.016 \pm0.034$ GeV$^{-2}$, significantly higher than 
$b^{\mathrm{GDH}}$ even within our quoted uncertainties. 
Note that $b^{\mathrm{GDH}}$ for the proton (neutron) alone is 0.456 GeV$^{-2}$ (0.518 GeV$^{-2}$), 
showing the delicate cancellation in the Bjorken integral that leads to this seemingly large deviation.
Rather than indicating a violation in the isovector sector of the GDH sum rule, a generic relation of quantum field theory,  
this deviation may reveal that $ \overline\Gamma_1^{p-n}(Q^2)$ has a quicker departure from the slope 
predicted by   the GDH sum rule than expected.
The tension could also possibly stem from the unmeasured low-$x$ contribution to $\overline\Gamma_1^{p-n}$. 
Although we have estimated  that contribution, it is difficult  to know its 
associated uncertainty because neither sufficient data nor firm theoretical guidance exist. Since many resonances that contribute to 
$\overline \Gamma_1^{p,n}$ cancel in $ \overline\Gamma_1^{p-n}$, notably the $\Delta$ resonances, the low-$x$ 
contribution  has {\it relatively} more weight in $ \overline\Gamma_1^{p-n}$ than in $\overline \Gamma_1^{p,n}$. 
In fact, fitting the measured part of $\overline \Gamma_1^{p-n}$ from EG4 before adding the estimated low-$x$ contribution yields 
$b^{\mathrm{no~low-}x}=0.093\pm0.032$ (see Table~\ref{table_fit3} in the Appendix), which shows
that a $100\%$ variation on the low-$x$ contribution would make $b$ from EG4 consistent with $b^{\mathrm{GDH}}$. 
The same finding holds with the EG4/E97110 data.  
Alternately, the finding that $b>b^{\mathrm{GDH}}$ could come from a systematic effect in the proton data 
since the EG4 and EG4/E97110 data sets partly share the same proton results.
However, the earlier $ \overline\Gamma_1^{p-n}$ data~\cite{Deur:2004ti} (open diamonds in Fig.~\ref{fig:bjsr}) 
already suggested the higher trend. 
Another possibility is that the extraction of $ \overline\Gamma_1^n$  from deuteron and 
$^3$He both have a systematic nuclear effect affecting them both in the same way, e.g. due to 2-body or 3-body break-ups or coherent contributions.


In conclusion, we presented new data on the Bjorken sum $\overline \Gamma_1^{p-n}(Q^2)$ in the 
$ 0.021 \leq Q^2 \leq 0.496$ GeV$^2$ range, which should cover well the domain of applicability 
of $\chi$EFT. 
The $\chi$EFT corrections to the leading order GDH  contribution are in the right direction and improve the agreement with the data significantly. 
 However, the agreement between the data and the two state-of-the-art $\chi$EFT curves is only marginal.
 In the case of~Ref.~\cite{Alarcon:2020icz},   the predictions of $\overline \Gamma_1^{p}$ 
and of $\overline \Gamma_1^{n}$ differ slightly from the respective data~\cite{Zheng:2021yrn, Sulkosky:2019zmn, Adhikari:2017wox}, with  these small differences not cancelling in $\overline \Gamma_1^{p-n}$. For Ref.~\cite{Bernard:2012hb}
the large differences observed above $Q^2 \approx 0.05$ GeV$^2$ between predictions and the $\overline \Gamma_1^{p}$ 
and $\overline \Gamma_1^{n}$ data do mostly cancel and
the $Q^2$ range over which the $\overline \Gamma_1^{p-n}$ data and prediction  display similar $Q^2$-behavior is much improved --by at least a factor of 3 to 5-- compared to 
$\overline \Gamma_1^{p}$, $\overline \Gamma_1^{n}$ and $\overline \Gamma_1^{p+n}$. In fact, the two $\chi$EFT predictions of $\overline \Gamma_1^{p-n}$ 
agree much better   with each other than for $\overline \Gamma_1^{p}$, $\overline \Gamma_1^{n}$ and $\overline \Gamma_1^{p+n}$, presumably because complications 
from their different treatment of the $\Delta$ resonance are largely absent. 
On the other hand, the $\Delta$ suppression makes accurate measurements of $\overline \Gamma_1^{p-n}$ challenging since it increases the relative importance of  the low-$x$  contribution compared to $\overline \Gamma_1^{p,n}$.
This may contribute to the tension between the data and the $\chi$EFT expectations. 
A future high-energy (up to $\nu=12$ GeV) measurement of the GDH sum at $Q^2=0$ on both the proton and the deuteron~\cite{Dalton:2020wdv}
will help constrain the low-$x$ contribution. 
Finally, our data, while in slight tension with the phenomenological models~\cite{Burkert:1992tg,Pasechnik:2010fg},  agree well with LFHQCD~\cite{Brodsky:2014yha}. 
Aside from testing non-perturbative descriptions of the strong force, the data   can be useful for extracting the QCD 
running coupling $\alpha_{g_1}$~\cite{Deur:2005cf} in the strong, yet near-conformal, regime of QCD.  

~

\begin{acknowledgments}
This material is based  upon work supported by the U.S. Department of Energy, Office of Science, 
Office of Nuclear Physics under contracts DE-AC05-06OR23177 and DE-FG02-96ER40960, and by the NSF under grant PHY-0099557.
  We gratefully acknowledge the contributions of the JLab CLAS and Hall A collaboration and JLab technical staff to the
preparation of the experiment, data taking and data analysis. 
\end{acknowledgments}

~

{\bf Appendix: Fit systematic studies}

\begin{table*}[t]
\begin{center}
\resizebox{\textwidth}{!}{%
\begin{tabular}{|c|c|c|c|c|c|} \hline
Data set                  &   $(a \pm uncor\pm cor)$           & $(b \pm uncor \pm cor )$ [GeV$^{-2}$]&$c\pm uncor \pm cor$ [GeV$^{-4}$] &$d\pm uncor \pm cor$ [GeV$^{-6}$]&  $\chi^2/n.d.f.$ \\ \hline 
EG4, no low-$x$   & NA                                              &$0.093\pm0.032 \pm0.000$     &   $-0.137\pm0.191 \pm0.000$                & NA &    1.24          \\  
EG4/E97110, no low-$x$   & NA                                &$0.112\pm0.022 \pm0.028$     &   $-0.123\pm0.118 \pm0.078$                & NA &    1.00          \\ \hline 
EG4                        & NA                                              &$0.170\pm0.032 \pm0.000$     &   $-0.046\pm0.191 \pm0.000$                & NA &    1.04          \\ 
EG4/E97110           & NA                                              &  $0.185\pm0.023 \pm0.027$     &   $-0.144\pm0.123 \pm0.075$              & NA  &    1.00          \\
{\bf World data}       & NA                                      &  $\bm{0.182\pm0.016 \pm0.034}$ &  $\bm{-0.117\pm0.091 \pm0.095}$ & NA &   {\bf 1.00}      \\ \hline 
World data             &   NA  &  $b^{\mathrm{GDH}}\equiv 0.0618$     &   $1.41\pm0.17 \pm0.39$      & $-4.30\pm0.80 \pm1.48$  &    1.97          \\  \hline 
World data             &   $(4.3\pm1.8 \pm0.1)\times 10^{-3}$   &  $0.092\pm0.042 \pm0.031$     &   $0.213\pm0.167 \pm0.086$      & NA  &    0.82          \\  \hline 
\end{tabular}}
\vspace{-0.2cm}
\caption{  
Fits of $ \overline\Gamma_1^{p-n}$ for different fit functions and data sets. 
The first column indicates the data set. 
Columns 2 to 4 give the values of the best fit coefficients and their uncertainties. (NA indicates that the term is not used in the fit.) 
The last column provides the $\chi^2/n.d.f.$. 
The first 5 rows are the best fits of $ \overline\Gamma_1^{p-n}$ using the fit function $bQ^2+cQ^4$.
The two first rows display, respectively, the results of the fits of the EG4 and EG4/E97110 data without unmeasured low-$x$ estimate. 
Rows 1 and 2 are to be compared to rows 3 and 4 that display the results of the fits of the full $ \overline\Gamma_1^{p-n}$ from the EG4 and EG4/E97110 data, respectively. 
The $5^{\mathrm{th}}$ row (bold fonts) indicates the best fit of the world data, {\it viz} the fit result reported in the main part of the article.
The $6^{\mathrm{th}}$ row displays the best fit of the world data using a $b^{\mathrm{GDH}}Q^2+cQ^4+dQ^6$ fit form, where $b^{\mathrm{GDH}}$
is set by the GDH sum rule prediction.
The last row shows the best fit of the world data for a $a+bQ^2+cQ^4$ fit form.
The fits are performed up to $Q^2 = 0.244$ GeV$^2$, the maximum
range of the E97110 data.}
\label{table_fit3}
\vspace{-0.7cm}
\end{center}
\end{table*}
To compare the data sets to each other and determine how well $b$ and $c$ 
are determined from the data, we performed fits over different subsets of the data. In addition, 
to assess the possible influences of higher order 
$Q^{2n}$-terms and of point-to-point correlated uncertainties, we also used fit functions  
allowing for a constant offset or for a higher order $Q^6$ term in the fit functions.
The most relevant fit results are provided in Table~\ref{table_fit3}. 
Furthermore, we also fit the EG4 and EG4/E97110 data without the unmeasured low-$x$ estimate down to
$x=0.001$ added to their integrals. Comparing the resulting fit parameters to the nominal ones ({\it viz} 
including the unmeasured low-$x$ estimate), permits us to assess the importance of the low-$x$ contribution. 
Table~\ref{table_fit3}  shows that in fact,  the low-$x$ contribution is sizable:  
45\% and 39\% of the values of   $b$ for the EG4 and EG4/E97-110 data, respectively.
%

The amount of systematic correlation between the data points being difficult to estimate,
we use the unbiased estimate method~\cite{Schmelling:1994pz, PDG2020}, 
where a fraction of the systematic uncertainty 
is added in quadrature to the statistical uncertainty for the total point-to-point uncorrelated uncertainty. 
This fraction is chosen such that the $\chi^2/n.d.f$ for our standard fit ($bQ^2 + c Q^4$) becomes 1 using the
total uncorrelated uncertainty of each data point (if the quadratic sum of the statistical and the entire 
systematic uncertainties is too small to reach $\chi^2/n.d.f.=1$,  then
$cor=0$ and $\chi^2/n.d.f.>1$). For the fit to the world data, this fraction is 58\%. 
The resulting uncertainty on the fit parameters is quoted as the ``uncor'' uncertainty in Tables~\ref{table_fit2} and \ref{table_fit3}. 
The  uncertainty ``$cor$'' was obtained by re-performing the fit with the central values of the data points 
(3\textsuperscript{rd} column of Table~\ref{table_data}) systematically shifted by the remaining 42\% 
of the systematic uncertainty.  The differences between the $b$ and $c$ hence obtained and the $b$ 
and $c$ from the nominal fit yield the ``$cor$'' uncertainties.
%

The fit results for the EG4 and EG4/E97110 data sets agree, irrespective of the chosen form of the fit.
Comparing the results in rows 1, 2 of Table~\ref{table_fit3}  to those of rows 3, 4 shows the large effect of the unmeasured low-$x$ contribution. 
The value for $c$ is consistent with zero for our main result, but depends strongly on the fit form. 
Like $b$, it is also strongly dependent on the low-$x$ contribution. 
While in most fits the central value of $c$ has the opposite sign to that 
predicted by $\chi$EFT, the signs agree if an offset $a$ is allowed
or if $b^{\mathrm{GDH}}$ is enforced. 
%
%

\end{document}